\documentclass[aps,prl,preprint,superscriptaddress,showpacs]{revtex4}

\usepackage{graphicx}
\usepackage{amsmath}
\usepackage{epsfig}

\begin{document}


\title{Weibel-instability-mediated collisionless shocks in laboratory with ultraintense lasers}

\author{F. Fiuza}
\email[Electronic address: frederico.fiuza@ist.utl.pt]{}
\affiliation{GoLP/Instituto de Plasmas e Fus\~ao Nuclear - Laborat\'orio Associado, Instituto Superior T\'ecnico, Lisboa, Portugal}

\author{R. A. Fonseca}
\affiliation{GoLP/Instituto de Plasmas e Fus\~ao Nuclear - Laborat\'orio Associado, Instituto Superior T\'ecnico, Lisboa, Portugal}
\affiliation{Instituto Universit\'ario de Lisboa, Lisboa, Portugal}

\author{J. Tonge}
\affiliation{Department of Physics \& Astronomy, University of California, Los Angeles, California 90095, USA}

\author{W. B. Mori}
\affiliation{Department of Physics \& Astronomy, University of California, Los Angeles, California 90095, USA}

\author{L. O. Silva}
\email[Electronic address: luis.silva@ist.utl.pt]{}
\affiliation{GoLP/Instituto de Plasmas e Fus\~ao Nuclear - Laborat\'orio Associado, Instituto Superior T\'ecnico, Lisboa, Portugal}

\begin{abstract}
The formation of non-relativistic collisionless shocks in laboratory with ultrahigh intensity lasers is studied via \emph{ab initio} multi-dimensional particle-in-cell simulations. The microphysics behind shock formation and dissipation, and the detailed shock structure are analyzed, illustrating that the Weibel instability plays a crucial role in the generation of strong subequipartition magnetic fields that isotropize the incoming flow and lead to the formation of a collisionless shock, similarly to what occurs in astrophysical scenarios. The possibility of generating such collisionless shocks in laboratory opens the way to the direct study of the physics associated with astrophysical shocks.
\end{abstract}



\maketitle

Understanding how collisionless shocks are formed and propagate in unmagnetized plasmas is of great importance to the study of many astrophysical scenarios such as gamma-ray bursts (GRB) afterglows, active galactic nuclei, pulsar wind nebulae, and supernova remnants \cite{bib:blandford1,bib:piran,bib:waxman}. The synchrotron radiation collected in astronomical observations suggests that these structures can generate subequipartition magnetic fields and accelerate particles to very high energies \cite{bib:jones}. How the magnetic fields are generated and what their structure is, which dissipation mechanism is dominant, which physical processes lead to shock formation, and how particles are accelerated, remain open questions. Since Coulomb scattering cannot be responsible for the dissipation process in a collisionless shock wave, anomalous heating associated with particle scattering in plasma turbulence seems to be the natural explanation \cite{bib:sagdeev}.

Electromagnetic turbulence associated with the Weibel, or current filamentation, instability \cite{bib:weibel} is believed to be the leading mechanism for shock formation in weakly magnetized plasmas \cite{bib:medvedev1}. This instability can generate small-scale magnetic fields in counterstreaming plasmas which can scatter particles and isotropize the flow. Previous numerical studies of idealized astrophysical collisionless shock scenarios have shown, using particle-in-cell (PIC) codes, that Weibel instability can lead to strong filamentation, magnetic field generation, and shock formation \cite{bib:simulations, bib:spitkovsky1} and that nonthermal particles can be accelerated in this shock structure \cite{bib:spitkovsky2,bib:martins1} and emit synchrotron radiation \cite{bib:sironi}. The validation of these numerical studies for astrophysical scenarios, where \emph{in situ} observations are not possible, is limited, since the information available from these astrophysical objects comes only from their radiation emission, which occurs at significantly larger temporal and spatial scales.

Laboratory experiments can play a crucial role in validating theoretical and numerical models of astrophysical phenomena \cite{bib:remington}. In the last years, a few experimental studies have been proposed and conducted for the generation of non-relativistic electrostatic collisionless shocks in laboratory with colliding laser-ablated plasmas \cite{bib:experiments}. However, in the case of Weibel mediated collisionless shocks, the conditions for shock generation in laboratory are not yet fully understood and no experimental evidence has been observed to date. The advent of high energy, high power laser systems is allowing for the exploration of extreme regimes in laser-plasma interactions, where strong plasma flows can be generated and the conditions necessary for the generation of Weibel mediated shocks in laboratory may be reached for the first time. Previous PIC studies of shock formation in intense laser-plasma interactions have focused on the one-dimensional (1D) dynamics \cite{bib:denavit,bib:silva2} and, therefore, could not evaluate the role/impact of Weibel instability in these scenarios.

In this Letter, we demonstrate the possiblity to generate truly Weibel mediated collisionless shocks in laboratory by the interaction of an ultraintense laser pulse with an overcritical plasma. Using \emph{ab initio} multi-dimensional relativistic PIC simulations, we examine in detail the physics behind shock formation and propagation, from the generation of the incoming flow by the intense laser to the microinstabilities that lead to the generation of subequipartition magnetic fields that isotropize the flow and lead to the formation of the shock structure. We show that the underlying physics is similar between these non-relativistic laser-driven shocks in laboratory and previously considered relativistic astrophysical shocks, illustrating the possibility of directly studying the physical mechanisms behind these astrophysical scenarios in laboratory.

In order to study the self-consistent shock formation and propagation in realistic laboratory conditions we use two-dimensional (2D) and three-dimensional (3D) simulations performed with OSIRIS \cite{bib:fonseca1}, a fully relativistic, electromagnetic, and massively parallel PIC code.
We simulate the interaction of an ultraintense laser pulse with a pre-ionized unmagnetized electron-proton plasma. The laser is linearly polarized and has a wavelength of 1 $\mu$m. We have simulated different laser intensities, ranging from $10^{20} - 10^{22}$ Wcm$^{-2}$, corresponding to a normalized laser vector potential $a_0 \sim 10-100$, and different plasma densities, ranging from $10 - 100 ~n_c$, where $n_c$ is the critical density for 1 $\mu$m light. The initial plasma temperature is 1 keV. The computational domain is typically 80 $c/\omega_{pi}$ in the longitudinal direction and 18 $c/\omega_{pi}$ in the transverse direction, with $c/\omega_{pi} = c (4\pi Z^2 e^2 n_p/m_i)^{-1/2}$ the ion skin depth for a plasma density $n_p$ and ion mass $m_i$; $e$ is the elementary charge, $Z$ is the charge state, and $c$ the speed of light in vacuum. The system is numerically resolved with 2 cells per $c/\omega_{pe}$ in both directions and uses 64 particles per cell for each species, for a typical total of $10^9$ particles. We use cubic particle shapes and current smoothing with compensation. Larger transverse box sizes, higher resolution, and higher number of particles per cell, were tested, showing overall result convergence. We note that these are the first full-scale simulations of unmagnetized electron-ion shocks (previous simulations used either positrons or ions with reduced mass ratios).

As the intense laser hits the overdense target it acts like a piston, pushing the front of the target as a massive and uniform flow but also generating a population of fast/hot electrons. Fast electrons, which typically have a density on the order of the critical density, $n_c$, and a relativistic factor $\gamma_0 \simeq \sqrt{1+a_0^2}$, are not affected by the proton response and propagate through the target. A cold return current is set up in order to balance the incoming fast electron flux. The two counterstreaming flows go Weibel unstable, similarly to what is believed to occur in astrophysics \cite{bib:medvedev1,bib:simulations,bib:spitkovsky1,bib:spitkovsky2,bib:martins1,bib:sironi}. Even in the case where the laser generated incoming flow is too hot to filament by itself in the background plasma, the return current is cold and therefore filaments. Following the usual procedure for the calculation of the dispersion relation for purely transverse modes \cite{bib:silva3}, and in the limit of a hot and rarefied electron flow counterstreaming with a cold and dense slowly drifting electron background, the maximum electron Weibel instability growth rate can be shown to be simply $\Gamma_{\mathrm{We}} \simeq (\beta_r/\sqrt{\gamma_r}) \omega_{pe}$, where $\beta_r$ is the normalized velocity of the returning electrons and $\gamma_r$ their Lorentz factor. In order to establish current neutrality $\beta_r \simeq n_c/n_p$, yielding $\Gamma_{\mathrm{We}} = 0.01 - 0.1 ~\omega_{pe}$ for the parameters in study, and leading to the generation of strong magnetic fields in a few $100 ~\omega_{pe}^{-1}$. The electron instability saturates when the magnetic energy density in the Weibel filaments becomes comparable to the energy density in the fast electron flow, leading to a saturation magnetic field amplitude $B_{sat} \simeq (8 \pi a_0 n_c m_e c^2)^{1/2}$, for ultrahigh laser intensities ($a_0 \gg 1$). The magnetic fields associated with Weibel/current filamentation instability of the fast electron flow isotropize the incoming non-relativistic electron-proton flow, leading to a strong compression and to the formation of a shock, defined as the density compression that propagates away from the laser-plasma interface. The shock speed is determined by the slowdown and mass/pressure build-up associated with this more massive flow, and therefore it is non-relativistic. Once the shock is formed, the particles that escape the shock from the downstream still provide the generation of the counterstreaming cold return current in the upstream, which allows for continuous filamentation in the upstream region. 

For the collisionless shock to be formed in the unmagnetized plasma, it is required that the piston (downstream) velocity exceeds the ion sound speed, $c_S = (Z k_b T_e / m_i)^{1/2}$, where $k_b$ is the Boltzmann constant, and $T_e$ is the bulk electron temperature, which is typically a fraction $\alpha \sim 1/3-1/2$ of the ponderomotive temperature before the shock is formed \cite{bib:josh}, $k_b T_e \simeq \alpha a_0 m_e c^2$. The downstream velocity, $v_d$, can be estimated by equating the momentum flux of the incoming mass flow with the laser light pressure, yielding a normalized velocity of $\beta_d = v_d/c = \sqrt{(n_c/2 n_p)(Z m_e/m_i) a_0^2}$ \cite{bib:wilks}. The condition for shock formation, $v_d > c_S$, is then given as a function of laser and plasma parameters by 
\begin{equation} 
a_0 \gtrsim 2 \alpha \frac{n_p}{n_c}.
\label{eq:shockform}
\end{equation}

\begin{figure}
\begin{center}
\includegraphics[width=0.7\textwidth]{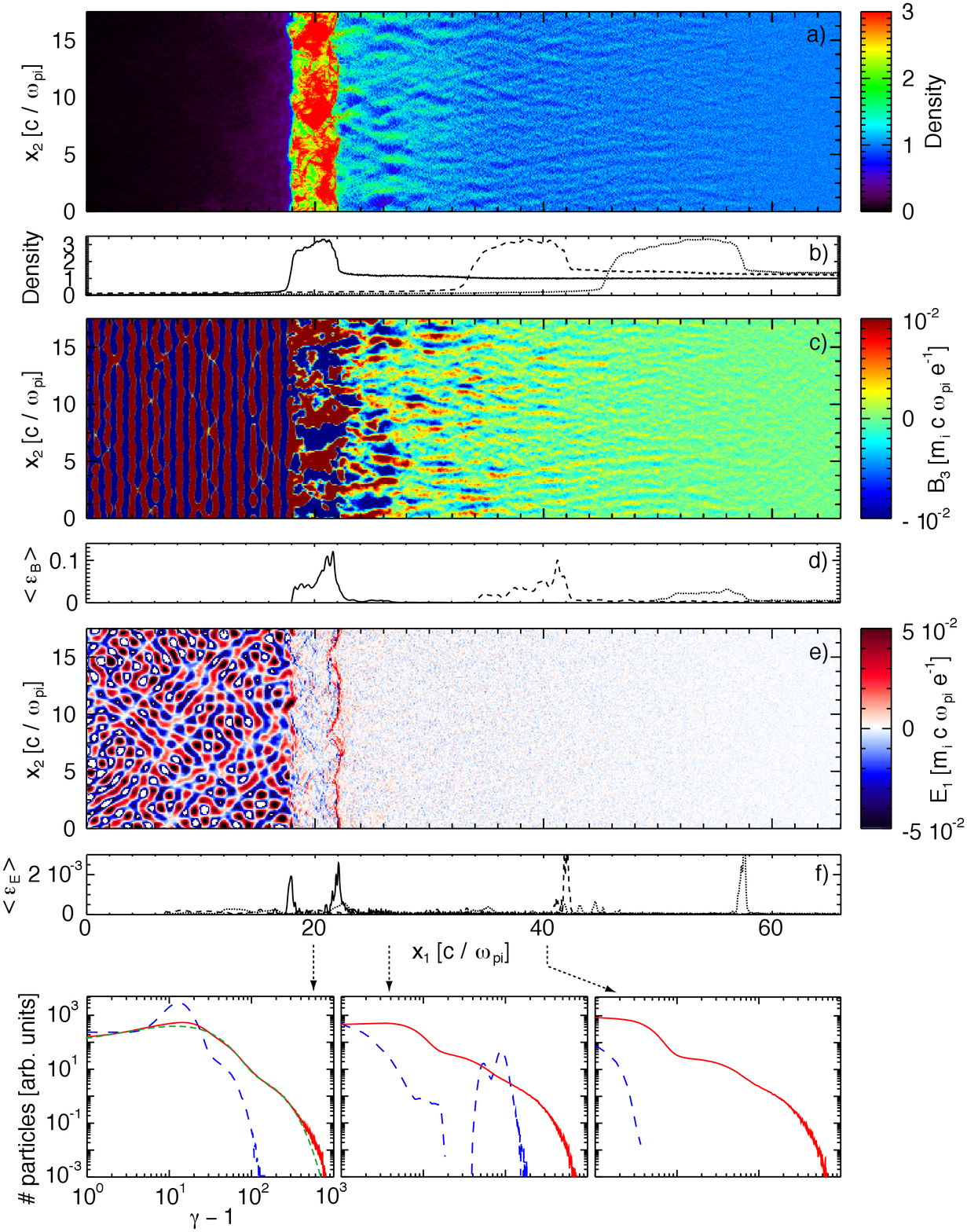}
\caption{\label{fig:shock} Steady state structure of a collisionless shock generated through the irradiation of an overcritical unmagnetized plasma by an ultraintense laser pulse after 385 fs (113 $\omega_{pi}^{-1}$) of interaction. a) Density structure normalized to the unperturbed upstream density. b) Transversely averaged plasma density. c) Magnetic field in the direction outside the simulation plane. d) Transversely averaged magnetic energy. e) Longitudinal electric field. f) Transversely averaged electric energy. g-i) Electron and ion (scaled up by the mass ratio $m_i/m_e$) spectrum at three different slices (positions marked by arrows). Red: electrons; blue: ions; green: fit to a sum of two 2D Maxwellians. The transversely averaged quantities in b), d), and f) are also shown for interaction times of 805 fs (250 $\omega_{pi}^{-1}$) (dashed) and 1134 fs (353 $\omega_{pi}^{-1}$) (dotted).}
\end{center}
\end{figure}

Fig. \ref{fig:shock} illustrates the main features of shock formation for a typical simulation where we have used a laser intensity of $5\times10^{21}$ Wcm$^{-2}$ ($a_0 = 60$) and a plasma density of $50 ~n_c$. A strong compression is observed in the downstream (Figs. \ref{fig:shock} a and \ref{fig:shock} b), between the laser-plasma interface and the shock front, and strong filamentation in the upstream region. The magnetic field illustrates similar filamentary structures (Fig. \ref{fig:shock} c). The filaments of density and magnetic field are not stationary in front of the shock. The filaments size evolves from the electron skin depth $c/\omega_{pe}$, far upstream, to the ion skin depth $c/\omega_{pi}$, close to the shock, and are then frozen behind the shock front. The shock transition is about 1-2 $c/\omega_{pi}$ thick, at early times, which is of the order of the ion Larmor radius, and corresponds to a peak in the magnetic energy (Fig. \ref{fig:shock} d). At later times, the thickness of the magnetic field peak continuously increases towards the downstream region, as observed in previous astrophysical configurations \cite{bib:keshet}, reaching values of the order of 10 $c/\omega_{pi}$ for our largest interaction times, 500 $\omega_{pi}^{-1}$. The transversely averaged magnetic field energy density reaches 12 \% of equipartition with the upstream kinetic energy density (measured in the downstream rest frame), \emph{i.e.} $\epsilon_B =  (B^2/8\pi)/(n_p m_i c^2 (\gamma_d - 1)) \simeq 0.12$ (Fig. \ref{fig:shock} d). Note that for non-relativistic shock velocities, the upstream velocity in the downstream frame is simply the opposite of the downstream velocity in the laboratory (upstream) frame, which in this case is measured to be $v_d \simeq 0.13 ~c$. Locally, the magnetic field energy density at the shock front can reach equipartition with the upstream, $\epsilon_{B,max} \simeq 1$. These values are fully consistent with previous simulations of Weibel mediated relativistic shocks in astrophysical scenarios \cite{bib:simulations,bib:spitkovsky1,bib:spitkovsky2,bib:martins1}, indicating that the underlying physical mechanisms are similar, even if in the present scenario the shock velocity is non-relativistic. An important difference is the well defined longitudinal electric field observed at the shock front in our simulations (Fig. \ref{fig:shock} e). This is associated with the fact that downstream electrons are significantly hotter than ions, since the laser predominantly heats electrons. This was not observed in previous simulations of relativistic counterstreaming plasmas, as both electron and ion flows are initialized completely cold, which will hardly be the case in a laser-driven laboratory configuration. Although the energy associated with this electric field is relatively small, $\epsilon_E =  (E^2/8\pi)/(n_p m_i c^2 (\gamma_d - 1)) \simeq 0.025$ (Fig. \ref{fig:shock} f), the field is able to reflect a fraction of the upstream ion population (10-15\%). As these reflected ions counterstream with the background plasma, they will lead to an enhancement of the magnetic fields in the ion time scales due to Weibel instability in the ions. In the limit of cold ions, the maximum ion Weibel growth rate is given by $\Gamma_{\mathrm{Wi}} \simeq \beta_b \sqrt{(n_b/n_p)/(\gamma_b m_i)} ~\omega_{pe} \simeq 2 \beta_{sh} \sqrt{(n_b/n_p)/m_i} ~\omega_{pe}$, where $n_b$ is the density of the reflected ion beam, which moves with twice the shock velocity. We can indeed see that in the foot of the shock (region where the reflected ions are present) the Weibel magnetic fields are stronger than in the remaining upstream region (Fig. \ref{fig:shock} c). At this point the instability becomes similar to more conventional scenarios with two counterstreaming plasma flows, but where the electrons are relativistic and ions are non-relativistic. It should be noted that an electrostatic ion-ion instability associated with the reflected ions has been previously identified in electrostatic shocks, strongly affecting its structure \cite{bib:kato}. It can be shown, following the usual procedure for the calculation of the dispersion relation for electrostatic modes \cite{bib:kato}, that in the limit of cold ion flows, the maximum growth rate of the ion-ion electrostatic instability is $\Gamma_{\mathrm{Ei}} \simeq \sqrt{(n_b/n_p)/(8 m_i)} ~\omega_{pe}$. This instability tends to dominate over the Weibel in the case of low shock velocities ($\beta_{sh} < 0.1$); however, for the large shock velocities reached in our proposed setup ($\beta_{sh} >0.1$) the Weibel instability dominates, further amplifying the magnetic fields and isotropizing the incoming flow. Even in the case where the initial maximum growth rates are comparable, the Weibel instability tends to dominate as the ion heating caused by the Weibel instability quickly shuts down the electrostatic instability. We have simulated the propagation of an ion flow in a plasma background to confirm this (not shown here). For the densities and velocities associated with our setup, and for different ion temperatures (1 eV - 1 keV), we consistently observe that the Weibel instability dominates over the electrostatic ion-ion instability, whereas in the case of lower ion velocities ($\beta_b < 0.2$) the electrostatic instability dominates. The condition $\beta_{sh} > 0.1$ effectively defines a lower limit for the laser intensity required to drive Weibel mediated shocks in this configuration. The detailed comparison of these instabilities in scenarios with high velocity ion flows will be given elsewhere.

The particle spectrum at different longitudinal positions is highly modified by the shock structure. In the downstream region (Fig. \ref{fig:shock} g), we observe a two-temperature electron spectrum from the laser acceleration, which can be reasonably fitted to a sum of 2D Maxwellians of the form $f(\gamma) = a_1 \gamma \exp(-\gamma/\Delta \gamma_1) + a_2 \gamma \exp(-\gamma/\Delta \gamma_2)$, with normalizations $a_1$ and $a_2$, $\Delta \gamma_1 = 13$, and $\Delta \gamma_2 = 58$, which is close to the expected laser induced ponderomotive temperature of $\Delta \gamma = \sqrt{1+a_0^2} \simeq 60$ \cite{bib:wilks}. The bulk electron temperature, $\Delta \gamma_1$, which initially is a fraction $\alpha$ of the ponderomotive temperature, changes as the shock is formed and most of the particles are trapped behind it, leading to an equipartition between the electron thermal energy and the ion fluid energy, \emph{i.e.} $\Delta \gamma_1 = (\gamma_d -1) m_i/m_e$. The most energetic electron population crosses the shock front into the upstream and remains relatively unchanged as evidenced by the spectra at the foot of the shock (Fig. \ref{fig:shock} h) and in front of it (Fig. \ref{fig:shock} i). The ions are heated up in the downstream (Fig. \ref{fig:shock} g), whereas in the upstream we observe a cold ion Maxwellian background and the presence, in the foot region, of a small ion population that has been reflected at the shock front (Fig. \ref{fig:shock} h). In order to compare directly our results to previous astrophysical simulations, we have transformed the different quantities into the downstream frame, in order to calculate the energy balance between electrons and ions at the different regions. As the filaments merge in the vicinity of the shock front and the magnetic fields of the filaments coalesce, the energy in the fields is converted back to thermal energy of the ions in the downstream region, thus leading to an effective heating of the ions and electrons. We observe that in the upstream region, the initial ion flow moving towards the shock loses 20-25\% of its energy for electron heating, and another 20-25\% goes into ion heating (thermal energy) in the downstream. These numbers are comparable, but lower than the observed values in idealized astrophysical configurations \cite{bib:astro}.

The observed features illustrate the collisionless nature of this shock structure, which has a transition region that is significantly smaller than the typical particle-particle scattering mean free path. The typical temperature and density behind the shock front is observed to be $T_i \sim 10 - 100$ keV, $T_e \sim 1-10$ MeV and $n_p \sim 30-300 n_c$, corresponding to a typical proton (electron) Coulomb mean free path on the order of $10^5 (10^6) ~c/\omega_{pi}$. In front of the shock the temperature can be significantly lower. In our simulations, with keV pre-ionized plasmas with densities of the order of a few tens of $n_c$ the proton (electron) Coulomb mean free path is still $\sim$ hundreds (few) $c/\omega_{pi}$, and therefore no significant collisional effects are expected. We redid our simulations with a relativistic Monte Carlo Coulomb collisional operator \cite{bib:fabio} in order to accurately model the effect of collisions and no significant differences have been observed. In this collisionless shock structure dissipation is provided by particle scattering in the self-generated Weibel turbulence. In other setups, where cold (10 - 100 eV), solid density ($> 100 n_c$) plasmas might be produced, $\nu_{ei}/\omega_{pe} \sim 0.1-1$ and therefore collisions between fast particles and the cold background plasma should be taken into account.

\begin{figure}
\begin{center}
\includegraphics[width=0.7\textwidth]{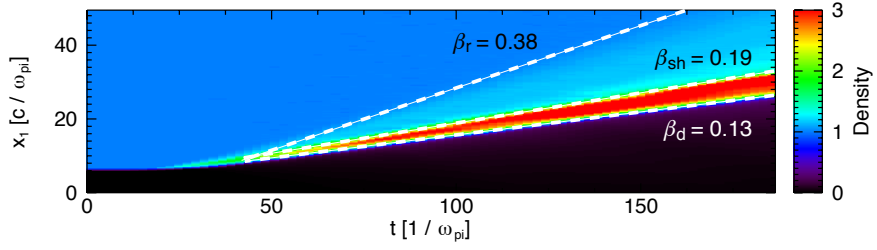}
\caption{\label{fig:xt} Temporal evolution of the transversely averaged density profile illustrating the normalized velocity of the downstream $\beta_d$, the shock $\beta_{sh}$, and the ions that are reflected at the shock front $\beta_r$.}
\end{center}
\end{figure}

The generated shock structure has a well defined velocity and density jump. The shock hydrodynamic jump conditions \cite{bib:blandford2} predict a density jump with a downstream to upstream density ratio $n_2/n_1 = (\Gamma_{ad} \gamma_d +1)/(\Gamma_{ad} -1)$ and normalized velocity of $\beta_{sh} \equiv v_{sh}/c = [(1+\Gamma_{ad} \gamma_d) (\gamma_d^2 -1)^{1/2})]/[1+\gamma_d + \Gamma_{ad} (\gamma_d^2 -1)]$, where $\Gamma_{ad}$ is the adiabatic index and $\gamma_d$ is the Lorentz factor of the downstream in the frame of the upstream (which for this case is the laboratory frame, since the upstream is approximately at rest). In the non-relativistic limit, where $\beta_d \ll 1$ and $\gamma_d \sim 1$, the density jump can be written as $n_2/n_1 = (\Gamma_{ad} +1)/(\Gamma_{ad} -1)$ and the shock velocity as
\begin{equation} 
\beta_{sh} \simeq \frac{a_0}{2} \sqrt{\frac{n_c}{2 n_p}\frac{Z m_e}{m_i}} (1+\Gamma_{ad}),
\label{eq:shockspeed}
\end{equation}
provided that the plasma is always opaque to the incoming light, \emph{i.e.} that the downstream density is higher than the relativistic critical density $n_p (\Gamma_{ad} +1)/(\Gamma_{ad} -1)  > n_c \sqrt{1+a_0^2}$. More general jump conditions, where the adiabatic index depends on the details of the distribution function of the downstream, can be used to infer the exact jump conditions corrected for distributions functions that deviate from Maxwellian and/or are different for the different plasma species and will be presented elsewhere \cite{bib:anne}. For the case of Fig. \ref{fig:shock}, we observe a density jump of $\sim 3.1$ and a normalized shock velocity of $\sim 0.19$ (Fig. \ref{fig:xt}), which are in very good agreement with the derived theoretical values of 3 and 0.2 respectively, for an adiabatic index of 2, appropriate for a 2D gas.

In order to understand if our 2D simulations can capture the relevant 3D physics, we have performed 3D simulations for the same parameters of Fig. \ref{fig:shock}. A similar shock structure has been obtained with a normalized shock velocity of 0.18, which is consistent with Eq. \eqref{eq:shockspeed} for an adiabatic index of 5/3, appropriate for a 3D gas. Conversely, 1D simulations of the same scenario cannot capture the relevant physics, in particular the current filamentation/Weibel instability, and therefore yielded different results.

\begin{figure}
\begin{center}
\includegraphics[width=0.7\textwidth]{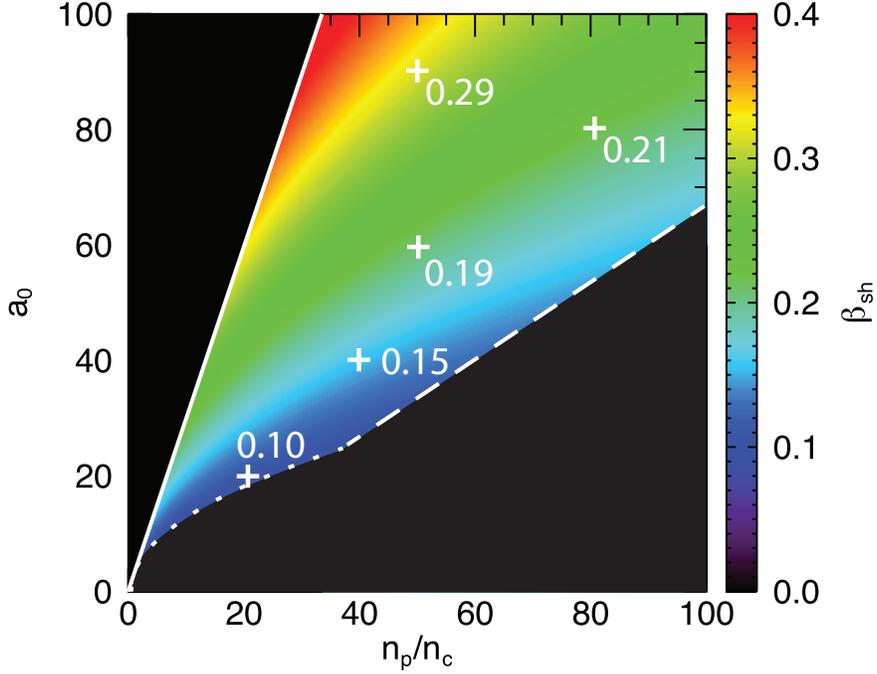}
\caption{\label{fig:shockmap} Shock velocity as a function of the plasma density and the laser normalized vector potential. Numerical values on the plot indicate the shock velocity measured in simulations. The parameter range has an upper limit defined by the condition for relativistic opacity of the downstream plasma (solid) and lower limits defined by the condition for the Weibel instability to dominate over the electrostatic instability, $\beta_{sh} > 0.1$ (dotted), and by Eq. \eqref{eq:shockform}, with $\alpha = 1/3$ (dashed).}
\end{center}
\end{figure}

As evidenced by the shock jump conditions, the properties of Weibel mediated collisionless shocks generated in laboratory in the configuration here proposed can be controlled by tuning the laser and plasma parameters. In particular the shock velocity can be controlled by adjusting the laser intensity, $I$, the plasma density, $n_p$, and/or the target composition (ion mass, $m_i$, and charge state, $Z$) according to Eq. \eqref{eq:shockspeed}, affecting proportionally the energy of the ions reflected by the shock, $\epsilon_{ion} [\mathrm{MeV}] \simeq 74.2 Z I[10^{21} \mathrm{Wcm}^{-2}]/n_p[10^{22} \mathrm{cm}^{-3}]$, which can be measured experimentally. This tunability is illustrated in Fig. \ref{fig:shockmap}, where we can see that the shock velocity obtained for different simulations with different laser and plasma parameters is in very good agreement with Eq. \eqref{eq:shockspeed}. Furthermore, the laser polarization can also be varied (from linear to circular) in order to control the laser electron heating and to study its influence in the shock properties.

The laser parameters required to study the generation of Weibel mediated collisionless shocks in laboratory can be estimated based on our results. The laser duration, $\tau_{0}$, should be significantly longer than the shock formation time, $t_f$, which is of the order of the time it takes the piston to push the plasma a shock thickness forward, $t_{f} \simeq (\beta_d \omega_{pi})^{-1}$,  \emph{i.e.} $\tau_0 \gtrsim 10 (\beta_d \omega_{pi})^{-1}$ or $\tau_0 [\mathrm{ps}] \gtrsim 0.5 \frac{m_i}{m_p}\left(I_0 [10^{21} \mathrm{Wcm}^{-2}] Z^3 \right)^{-1/2}$, where $m_p$ is the proton mass. The laser spot size should be much larger than the shock thickness in order to have a stable shock front, $W_0 \gtrsim 10 c/\omega_{pi}$ or $W_0 [\mu\mathrm{m}] \gtrsim 10 \left(\frac{m_i}{m_p}\right)^{1/2} \frac{\lambda_0 [\mu\mathrm{m}]}{Z} \left(\frac{n_p}{50 n_c} \right)^{-1/2}$.
Combining the two previous conditions, the laser energy required to provide a stable shock structure is $\epsilon_0 [\mathrm{kJ}] \simeq 1.5 \left(\frac{I_0 [10^{21} \mathrm{Wcm}^{-2}]}{Z^7} \right)^{1/2} \left(\frac{m_i}{m_p}\right)^{2} \left(\lambda_0 [\mu\mathrm{m}]\right)^2 \left(\frac{n_p}{50 n_c} \right)^{-1}$. We note that high-Z materials allow the use of lower energy lasers in order to drive a stable shock structure. Current and near-future laser systems can deliver the required picosecond 100 J - kJ pulses, either directly \cite{bib:lasers} or using a recently proposed Raman amplification scheme to convert nanosecond kJ pulses into picosecond kJ pulses \cite{bib:trines}. We also note that the simulations here shown are valid for different laser wavelengths and plasma densities, provided that the laser to plasma frequency ratio is kept constant and that collisional effects can still be neglected. For instance, CO2 lasers interacting with gas jets, which can easily provide densities above the critical density for $10 ~ \mu$m laser wavelength, could be used in order to generate similar shock structures \cite{bib:haberberger}.

The versatility of the setup here presented should also allow for the generation of laboratory shocks in different astrophysical relevant regimes. Namely, the use of non-uniform targets would allow for the study of shock formation and propagation in a clumpy medium where magnetic energy production by macroscopic turbulence may become important \cite{bib:sironi2}, and the use of strong, externally induced, uniform magnetic fields \cite{bib:krauer} can be envisioned to study shock formation and propagation in plasmas with variable magnetization.

In conclusion, we have studied the generation of Weibel mediated collisionless shocks in unmagnetized plasmas, driven by the interaction of an ultraintense laser pulse with an overcritical plasma. We have shown that in realistic laboratory conditions the plasma flow generated by this interaction can lead to the generation of subequipartition magnetic fields, due to Weibel/current filamentation instability, that isotropize the flow and generate a shock. The shock structure and its properties,
here shown for the first time for realistic ion to electron mass ratios, are similar to previously simulated low mass ratio Weibel mediated collisionless shocks in idealized astrophysical scenarios. Our results illustrate the possibility of studying for the first time in laboratory the physics behind the formation and propagation of Weibel mediated collisionless shocks in unmagnetized plasmas, which would allow for a better understanding of the role of these structures in nonthermal particle acceleration and emission of synchrotron radiation in astrophysical scenarios.

Work supported by the European Research Council (ERC-2010-AdG Grant 267841) and FCT (Portugal) grants PTDC/FIS/111720/2009 and SFRH/BD/38952/2007. We would like to acknowledge the assistance of high performance computing resources (Tier-0) provided by PRACE on Jugene based in Germany. Simulations were performed at the IST cluster (Lisbon, Portugal), the Hoffman cluster (UCLA), and the Jugene supercomputer (Germany).


\newpage

\end{document}